# Unprecedented High Irreversibility Line in Nontoxic Cuprate Superconductor $(Cu,C)Ba_2Ca_3Cu_4O_{11+\delta}$

Yue Zhang[1], Wenhao Liu[1], Xiyu Zhu[1,2], Haonan Zhao[1], Zheng Hu[1], Chengping He[1] & Hai-Hu Wen[1,2]*

[1] National Laboratory of Solid State Microstructures and Department of Physics, Nanjing University, Nanjing 210093, China

[2] Collaborative Innovation Center of Advanced Microstructures, Nanjing University, Nanjing, 210093, China

*Corresponding author. Email: hhwen@nju.edu.cn (H.H.W.)

**Abstract**: **One of the key factors limiting the high power applications for a type-II superconductor is the irreversibility line $H_{irr}(T)$ which reflects the very boundary of resistive dissipation in the phase diagram of magnetic field versus temperature. In cuprate family, the Y-, Bi-, Hg- and Tl-based systems have superconducting transition temperatures exceeding the liquid nitrogen boiling temperature (~77K). However, the toxic elements Hg and Tl in the latter two systems strongly constrain the possible applications. The Bi-based (2223) system is nontoxic, but the irreversibility magnetic field is strongly suppressed in the liquid nitrogen temperature region. For this purpose, the best perspective so far is relying on the $YBa_2Cu_3O_{7-\delta}$ ($T_c \approx 90$ K) system which is nontoxic and has a relatively high irreversibility magnetic field. Here we report the study of a nontoxic superconductor $(Cu,C)Ba_2Ca_3Cu_4O_{11+\delta}$ with $T_c$ = 116K. It is found that *the irreversibility magnetic field is unprecedentedly high among all superconductors* and it thus provides a great potential of applications in the liquid nitrogen temperature region.**

**One Sentence Summary:** The nontoxic superconductor $(Cu,C)Ba_2Ca_3Cu_4O_{11+\delta}$ has the highest irreversibility line among all superconductors, yielding a promising high power application in the temperature region of 100 K.

For a type-II superconductor, the magnetic field will penetrate into the sample and form the quantized vortices when the external magnetic field is beyond a threshold $H_{c1}$. This state, with the mixture of superconducting area and magnetic vortices, is called as the mixed state. The upper boundary of the mixed state is the upper critical magnetic field $H_{c2}$ at which the Cooper pairs are completely absent (BCS picture). However, for a superconductor to carry the non-dissipative supercurrent, another boundary, namely the irreversibility line $H_{irr}(T)$ is very crucial. This phase line actually separates the phase diagram into zero and finite resistance. In the cuprate family, the Bi-based (*1*), Hg-based (*2*) and Tl-based (*3*) systems show superconducting transition temperatures beyond 100K. Among all these materials, it was reported that the $(Tl_{0.5}Pb_{0.5})Sr_2Ca_2Cu_3O_9$ (*4, 5*) exhibits a rather high irreversibility line with an irreversibility field of nearly 10T at the liquid nitrogen boiling point (~77K), which is thought to be hopeful for applications. However, for Hg-based and Tl-based systems, the toxic elements Hg and Tl strongly limit the application of these materials. The nontoxic Bi-based (2223, $T_c\approx$110K) system also has a transition temperature exceeding 100K, but the very layered structure does not allow a high irreversibility field at the liquid nitrogen temperature (*6, 7*), the irreversibility field and superconducting current density decreases rapidly with increasing temperature at liquid nitrogen point (~77K). For the nontoxic $Bi_2Sr_2CaCu_2O_8$ system, the material can be made into round wire and the irreversibility field is high at the liquid He temperature, therefore it could lead to the high magnetic field application at low temperatures (*8*). The Y-based $YBa_2Cu_3O_{7-\delta}$ ($T_c\approx$90 K), which is nontoxic and has a high irreversibility field (about 8-10T at 77K) (*9, 10*), is thought to be a promising material for applications. By irradiation, the first order melting of vortex lattice in YBCO will be turned into second order, but the irreversibility line does not change too much (*11*). In this paper, we present systematic study on the synthesis, measurements of resistivity and magnetization in nontoxic superconductors $(Cu,C)Ba_2Ca_3Cu_4O_{11+\delta}$ with $T_c$ = 116K, which exhibit the highest irreversibility line so far among all superconductors in the liquid nitrogen temperature region. It is found that the irreversibility field $\mu_0H_{irr}$ of $(Cu,C)Ba_2Ca_3Cu_4O_{11+\delta}$ we made in this work reaches about 15T at 86K, and 5T at 98K, providing a promising high power application in the temperature region of 100 K.

The $(Cu,C)Ba_2Ca_3Cu_4O_{11+\delta}$ bulk samples were made by solid state reaction method under high pressure and high temperature. The details of synthesis and characterization are given in Supplementary Materials. The resistivity and magnetization measurements were conducted with the Quantum Design instruments PPMS16T and SQUID-VSM7T, respectively. The powder x-ray diffraction (XRD) pattern of Sample 1 can be fitted quite well with the standard (Cu,C)-1234 XRD pattern (fig. S4). It shows a dominant phase of (Cu,C)-1234 tetragonal phase with a space group P4/mmm (the fitting fraction is about 93%) with parameters a = 3.86Å and c = 17.94 Å. The main impurity phase is $CaCuO_2$ (< 7%) which is marked with the green vertical lines below the spectrum. Two stars indicate the unknown phase which should take a very small fraction of the sample.

We measured the temperature dependent resistivity of Sample 1 under different magnetic fields (Fig. 1A). One can see that the superconducting transition occurs at about 116 K with a rather narrow transition width ($\Delta T_c \approx$1K) determined from the 10-90% normal state resistivity. In the inset of Fig. 1A, we show the temperature dependence of magnetic susceptibility measured in zero-field-cooled (ZFC) and field-cooled (FC) modes for the as-synthesized Sample 1 $(Cu,C)Ba_2Ca_3Cu_4O_{12}$ under 10 Oe. A calculation of the magnetic screening volume using the ZFC data at 10 Oe is about 196%. The value of being larger than 100% is due to the demagnetization factor. Again the diamagnetic susceptibility measurement shows a well-defined transition at about 116K, being consistent with the resistivity measurements.

The superconducting transition is getting broad under a magnetic field (Fig. 1, A and B), which is induced by the vortex motion. Since our sample is a bulk one, the crystallographic direction is random in different grains. We would say that the threshold of dissipation is determined by the part with magnetic field parallel to c-axis in some grains or the weak link areas. As we can see, the magnetic field induced broadening is rather slow compared to the nontoxic Bi-based 2223 system. For example, if we adopt the criterion of resistivity of 1%$\rho_n$ ($T_c$), as marked by the blue horizontal dashed line, the irreversibility field is 15T at about 82 K (Fig. 1B). With the same criterion, we have an irreversibility field of about 3T at 98.6 K, and 5T at 94.2K. Actually we have found even higher values in another sample (Sample 2). The irreversibility fields are 15T at 86K, 5T at 98K and 3T at 101.2K (fig. S1). All these values suggest that the nontoxic system $(Cu,C)Ba_2Ca_3Cu_4O_{11+\delta}$ has the highest irreversibility line among all superconductors in the liquid $N_2$ temperature region.

In order to evaluate the intrinsic critical current density and their behavior under magnetic field, we crash the sample into powder of grains with the averaged size of about 5 μm and measured the magnetization. The averaged grain size is measured by scanning electron microscope (SEM) (fig. S3). Here we use the hydrochloric acid to corrode the surface of bulk Sample 1 for a few seconds in order to clearly show the grains. We also measured the magnetization hysteresis loops (MHLs) of the Sample 1 in powder state from 4.2K to 110K for applied fields up to 7T (Fig. 2A). One can see that the magnetization is irreversible up to quite high temperature and magnetic field. For the powdered superconducting sample, the critical current density $J_c$ is determined using the Bean critical-state model $J_c$ (A/cm$^2$) = 30ΔM/d (*12-14*), where ΔM (in units of emu/cm$^3$) is the total magnetization, d (in units of cm) is the averaged size of the grains. Using Bean critical-state model, we obtain the critical current density $J_c$ vs magnetic field at different temperatures (Fig. 2B). It is clear that the $J_c$ value can reach about $6\times10^6$ A/cm$^2$ at 4.2 K at the earth field. And the critical current is quite robust against the magnetic field. Using a criterion of 1000A/cm$^2$, we get an irreversibility field of about 2 Tesla at 100K, which is slightly lower than that determined from the resistive measurements because the criterion of $J_c$ adopted here is rather high. At about 90 K, the critical current density $J_c$ keeps flat versus magnetic field up to 7 Tesla, indicating a much higher irreversibility magnetic field.

As shown above, the critical current density and the irreversibility line in the $(Cu,C)Ba_2Ca_3Cu_4O_{11+\delta}$ system are both very high. Here we would like to show a comparison of the irreversibility lines in different cuprate systems. Worthy of mentioning is that we obtain the irreversibility line of Sample 1 using a resistivity criterion of 1%$\rho_n$ ($\rho_n \approx 0.09$ mΩ·cm), as shown by the dashed blue line (Fig. 1B), which is also adopted by other groups. Fig. 3 shows the comparison of irreversibility lines for (Cu,C)-1234(Sample 1 and Sample 2 in this work), YBCO(single crystals), Bi-2223(thin film), Bi-2223(single crystal) and (Tl,Pb)-1223. It is clear that the nontoxic Bi-based (2223) has a very low irreversibility field at 77 K, while YBCO has a relatively high $\mu_0 H_{irr}$ (about 8-10T at 77K, H||c-axis) compared with Bi-2223. The material (Tl,Pb)-1223 shows an higher $T_c$ and

irreversibility line (*4, 5*) than YBCO while it has toxic element Tl. However, the irreversibility field of the nontoxic samples (Cu,C)-1234 we synthesized show much steeper temperature dependence. In previous studies, the magnetization were measured for the same system (Cu,C)-1234 (*13*) or Au-1234 (*14*), but the irreversibility lines reported there are not as high as our samples. This may be induced by the difference of the material morphologies inside the sample. As we mentioned already, there is an irreversibility field of 15T at temperature of 82K for Sample 1; it even shoots up to 86K under 15T in Sample 2. Here we use the highlighted area to indicate the region with finite supercurrent (or zero/weak resistive dissipation) compared with YBCO (Fig. 3). One can see that there is a large area beyond the irreversibility line of YBCO where the samples can carry non-dissipative supercurrent, providing great potential of application in the temperature region of 100 K. We must emphasize that the present samples were made through the high pressure synthesis. Our present results just show very good intrinsic properties for applications of the non-toxic material (Cu,C)-1234. It is highly desired to try new method with lower pressure or thin film deposition to make the superconducting wire/tape based on this promising material.

**Acknowledgments:** We acknowledge the useful discussions with W. K. Kwok; **Funding:** This work was supported by National Key R&D Program of China (grant no. 2016YFA0300401), National Natural Science Foundation of China (grant nos. 11534005); **Author contributions:** The sample growth was done mainly by Y. Z. with the help of W. L., X. Z., H. Z., Z. H., C. H.. The resistivity and magnetization were done and analyzed by Y. Z. and H.-H.W. The XRD was measured and analyzed by Y. Z. and X. Z. H.-H.W and Y. Z contributed to the writing of the paper. H.H.W. coordinated the whole work. All authors have discussed the results and the interpretations. **Competing interests:** The authors declare no competing financial interests. **Data and materials availability:** All data needed to evaluate the conclusions of the paper are present in the paper and/or the Supplementary Materials. Additional data related to this paper may be requested from the authors.

## Figures and captions

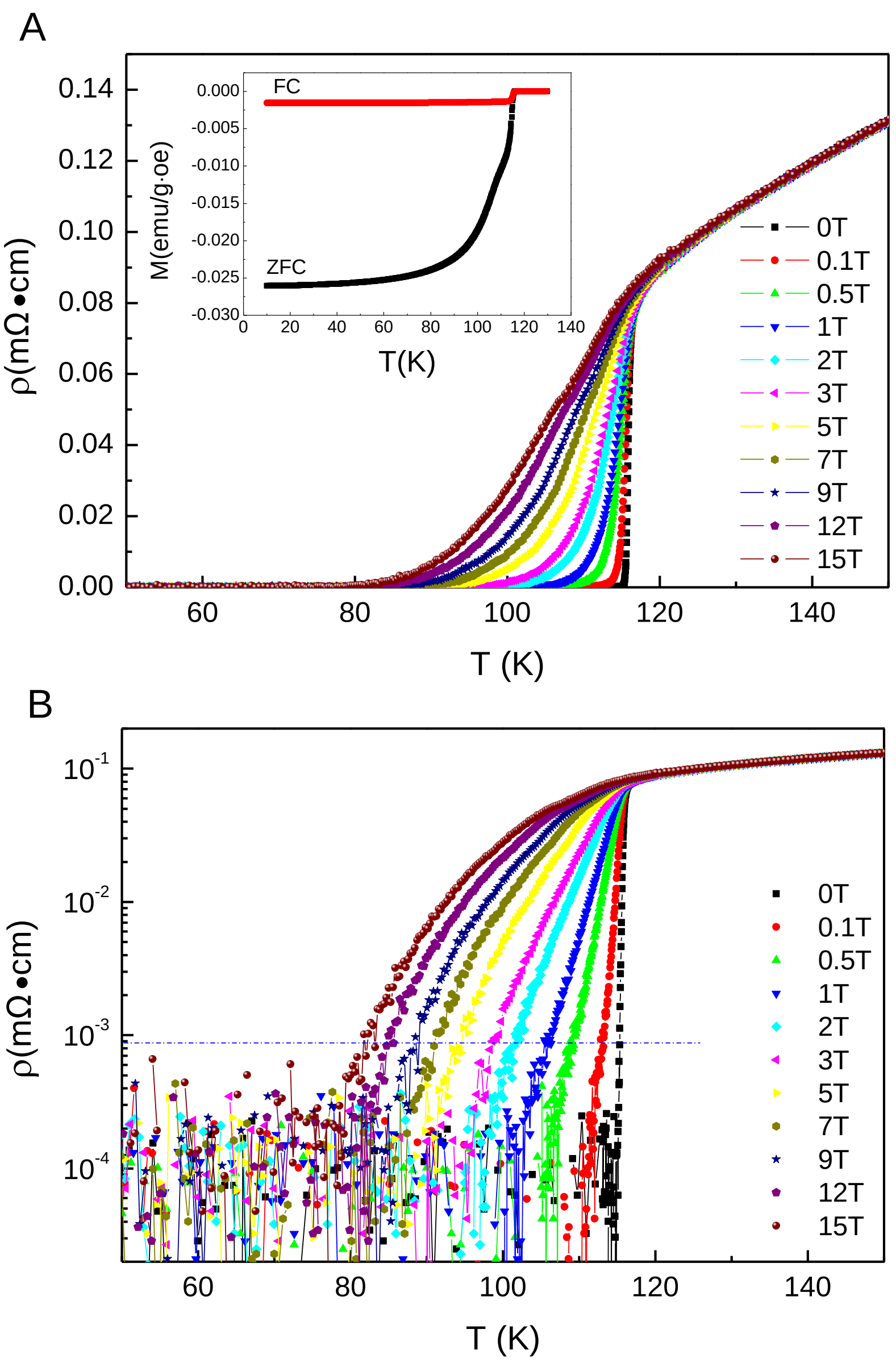

**Fig. 1.** Temperature dependence of resistivity and magnetization of Sample 1. (**A**) Temperature dependence of resistivity under different magnetic fields from 0 to 15 T. The inset shows the temperature dependence of magnetic susceptibility measured in the ZFC and FC modes under a magnetic field of 10 Oe. (**B**) The same data in the main panel of A in the semi-logarithmic scale. The blue horizontal dashed line represents the criterion of resistivity $0.01\rho_n(T_c)$ which is used to determine the irreversibility line.

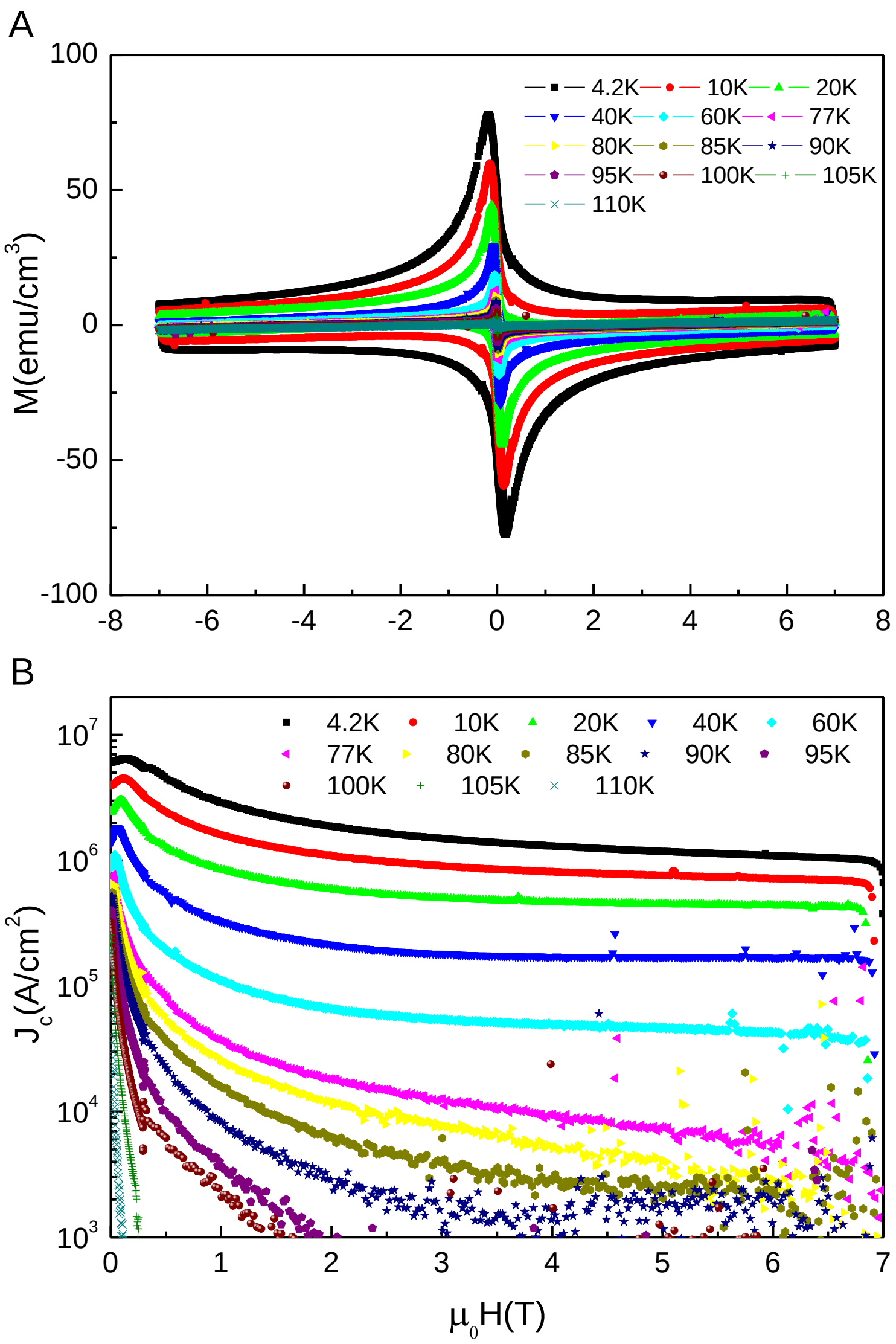

**Fig. 2.** Magnetization and critical current density of powdered sample (Sample 1). (**A**) Magnetization hysteresis loops (MHLs) of high pressure synthesized (Cu, C)-1234 from 4.2K to 110K under magnetic fields up to 7T. (**B**) Field dependence of critical current density $J_c$ at various temperatures. $J_c$ is determined by the formula $J_c$ (A/cm$^2$) =30ΔM/d, where d (in units of cm) is the averaged grain size of the powder sample. Some strongly scattered data are the bad points during the SQUID measurements on powdered samples. Due to the presence of an unphysical hysteresis background in our SQUID-VSM measurements corresponding to a small current density, we have deducted the background logJc=0.295H+1.84 from the data, with $J_c$ in units of A/cm$^2$ and H in tesla.

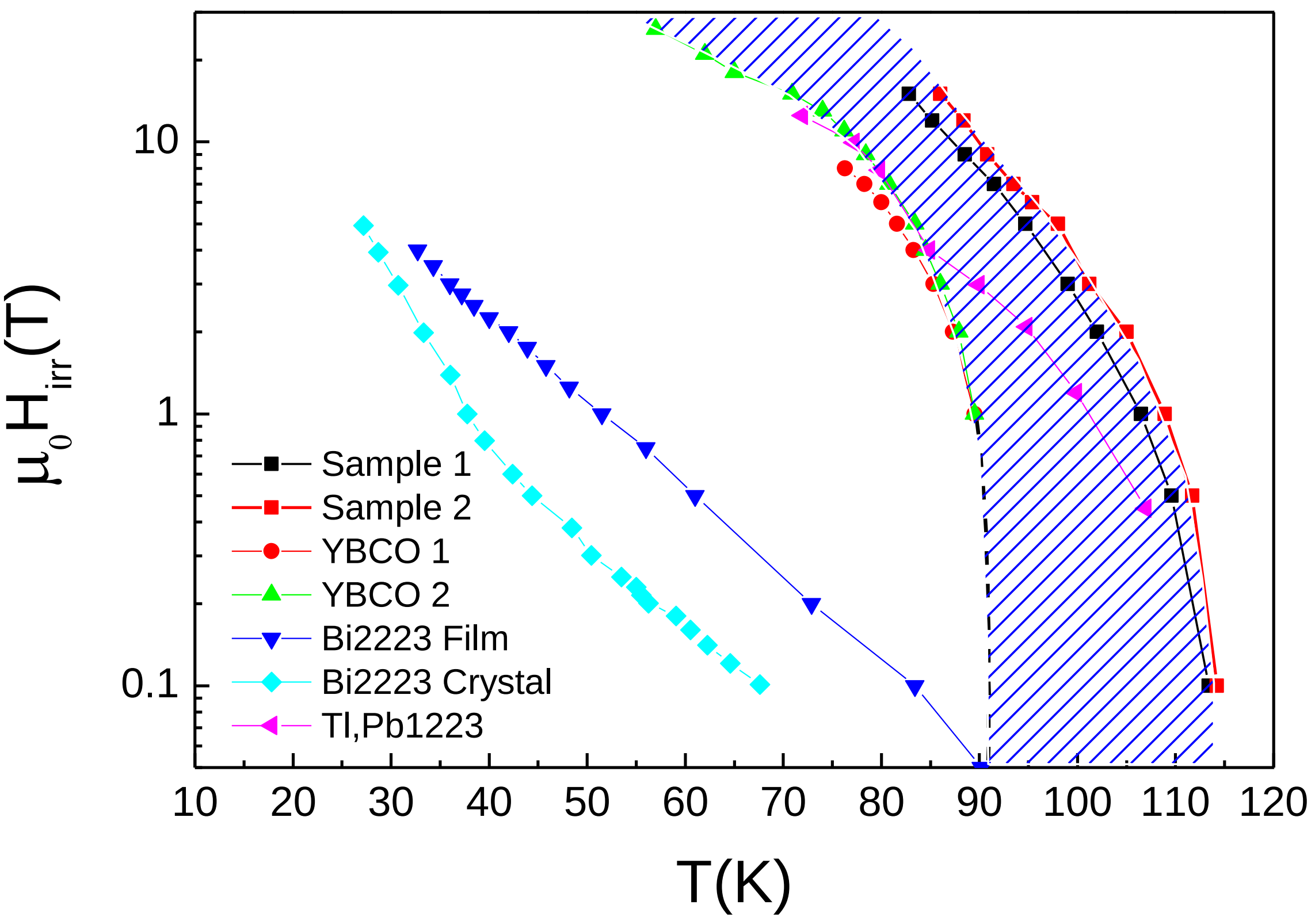

**Fig. 3.** Irreversibility lines of different cuprate systems. Irreversibility lines for (Cu,C)-1234(this work, Sample 1 and Sample 2) , YBCO 1 and 2 (single crystals, H||c-axis), Bi-2223(thin film, blue down triangles), Bi-2223(single crystal, Cyan diamond) and (Tl,Pb)-1223 (pink triangles). The highlighted area indicates the region for zero dissipation above the boundary of YBCO. The dark dashed line shows the trend of irreversibility line of YBCO with $T_c$=91K.

## Supplementary Materials

# Unprecedented High Irreversibility Line in Nontoxic Cuprate Superconductor $(Cu,C)Ba_2Ca_3Cu_4O_{11+\delta}$

Yue Zhang[1], Wenhao Liu[1], Xiyu Zhu[1,2], Haonan Zhao[1], Zheng Hu[1], Chengping He[1] & Hai-Hu Wen[1,2]*

### Materials and Methods

**Sample synthesis under high pressure and high temperature.** The bulk samples were made by solid state reaction method under high pressure and high temperature using the appropriate precursors $BaCuO_{2.13}$ and $Ca_2CuO_3$. Briefly, $BaCuO_{2.13}$ was obtained by calcining $BaO_2$ and CuO at 900°C in flowing oxygen gas for a total duration of 60h with two intermediate grindings for powder homogeneity. The precursor $Ca_2CuO_3$ was prepared first by calcining a well-ground mixture of $CaCO_3$ and CuO at 950°C in air for 20h, then the obtained $Ca_2CuO_3$ powder was sintered in flowing oxygen gas at 950°C for 40h with one or two intermediate grinding. After grinding $BaCuO_{2.13}$, $Ca_2CuO_3$, $CaCO_3$, $BaCO_3$, CuO and appropriate amount of $Ag_2O$ (act as oxidizer), the mixture was pressed into pellets which were sealed into a gold capsule for the final step of high pressure and high temperature synthesis. For high pressure synthesis, the sample enclosed by gold capsule was placed into a BN container, surrounded by graphite sleeve resistance heater and pressure transmitting MgO rods. The final reaction was carried out at 1100-1150°C under 3.5GPa for 1-3h, then cooled down to room temperature in 5 min before the pressure was released.

**Resistivity and magnetization measurement.** Resistivity, magnetization and x-ray diffraction (XRD) were measured systematically on the samples. The dc magnetic susceptibility was measured by a Quantum Design instrument SQUID-VSM7T. Magnetization hysteresis loops measurement was performed with magnetic fields up to 7T. We also measured the resistivity curves versus temperature with magnetic field from 0 to 15T by the standard four-probe method using a physical property measurement system (PPMS16T, Quantum Design). The x-ray diffraction (XRD) was measured using a Bruker D8 Advanced diffractometer with the CuK$\alpha_1$ radiation at room temperature.

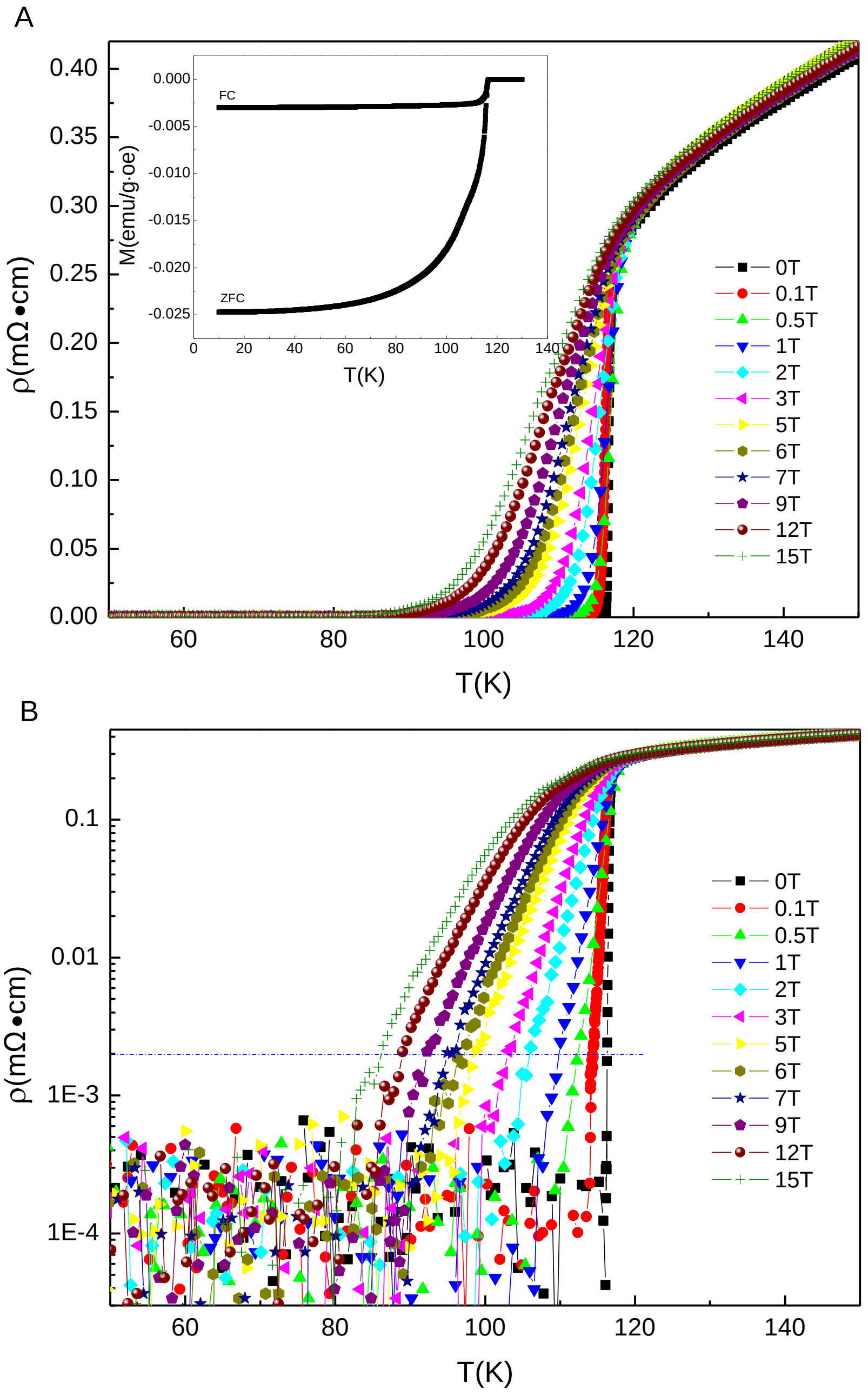

**Figure S1.** Resistivity and magnetization versus temperature of Sample 2. (**A**) Temperature dependent resistivity under different magnetic fields from 0 to 15T. Magnetic susceptibility vs temperature under a magnetic field of 10Oe is shown in the inset of Fig. 1A. (**B**) The same data of Sample 2 in the main panel of A in the semi-logarithmic scale. The blue horizontal

dashed line represents the criterion of resistivity $0.01\rho_n(T_c)$ which is used to determine the irreversibility line.

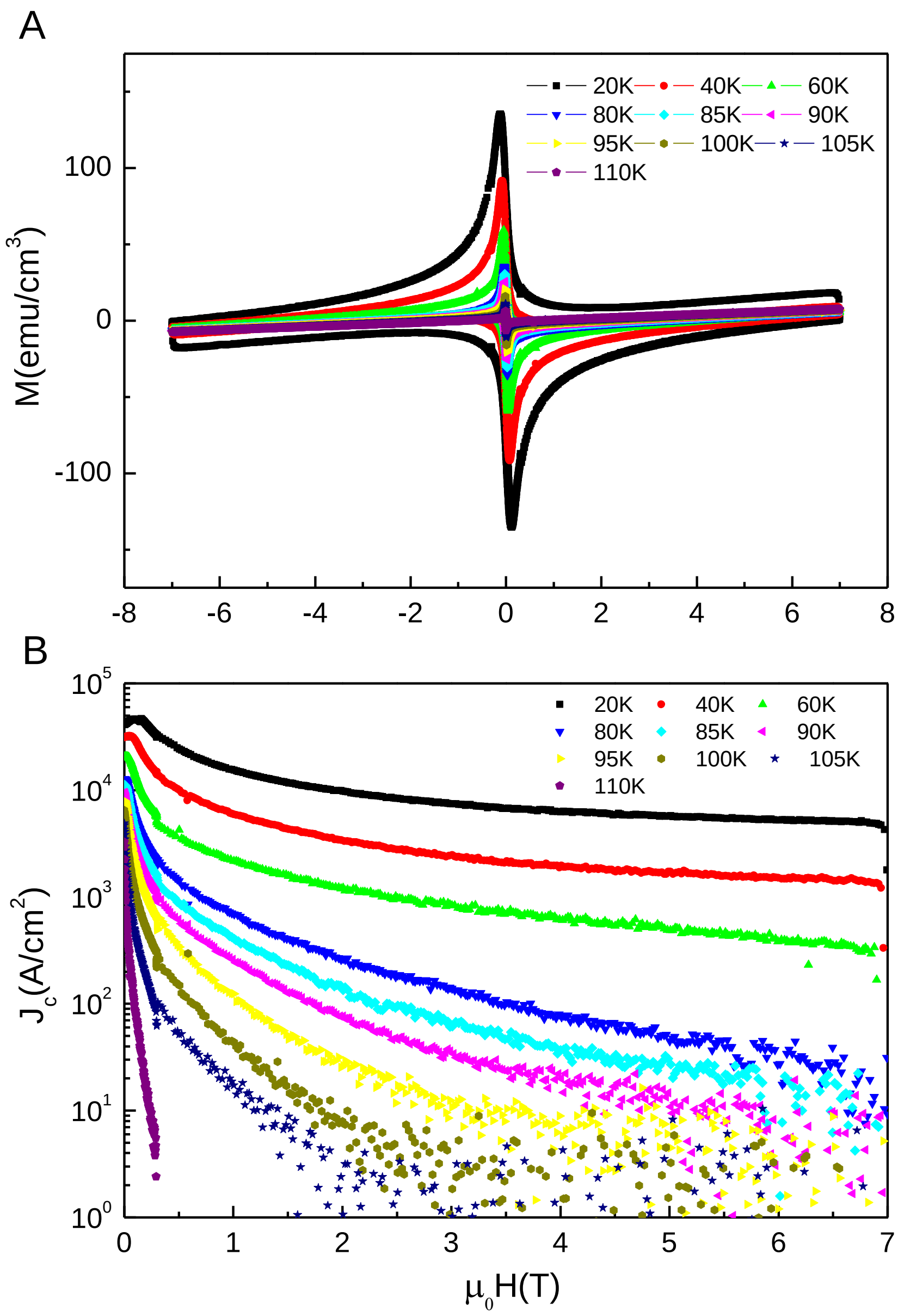

**Figure S2.** Magnetization and critical current density of bulk Sample 2. (**A**) MHLs of bulk Sample 2 from 20K to 110K under magnetic fields up to 7T. (**B**) Magnetic field dependence of critical current density $J_c$ at various temperatures of Sample 2. $J_c$ is determined by Bean's critical state model $J_c$ (A/cm$^2$) =20ΔM/w(1-w/3$l$), where $l$ (in units of cm) is the length of the sample, while w (in units of cm) is the width of the sample. Unfortunately for this sample we did not measure the MHLs for powdered sample. Here an unphysical background logJc=0.27H+0.017 due to the SQUID-VSM problem in the small current region is deducted.

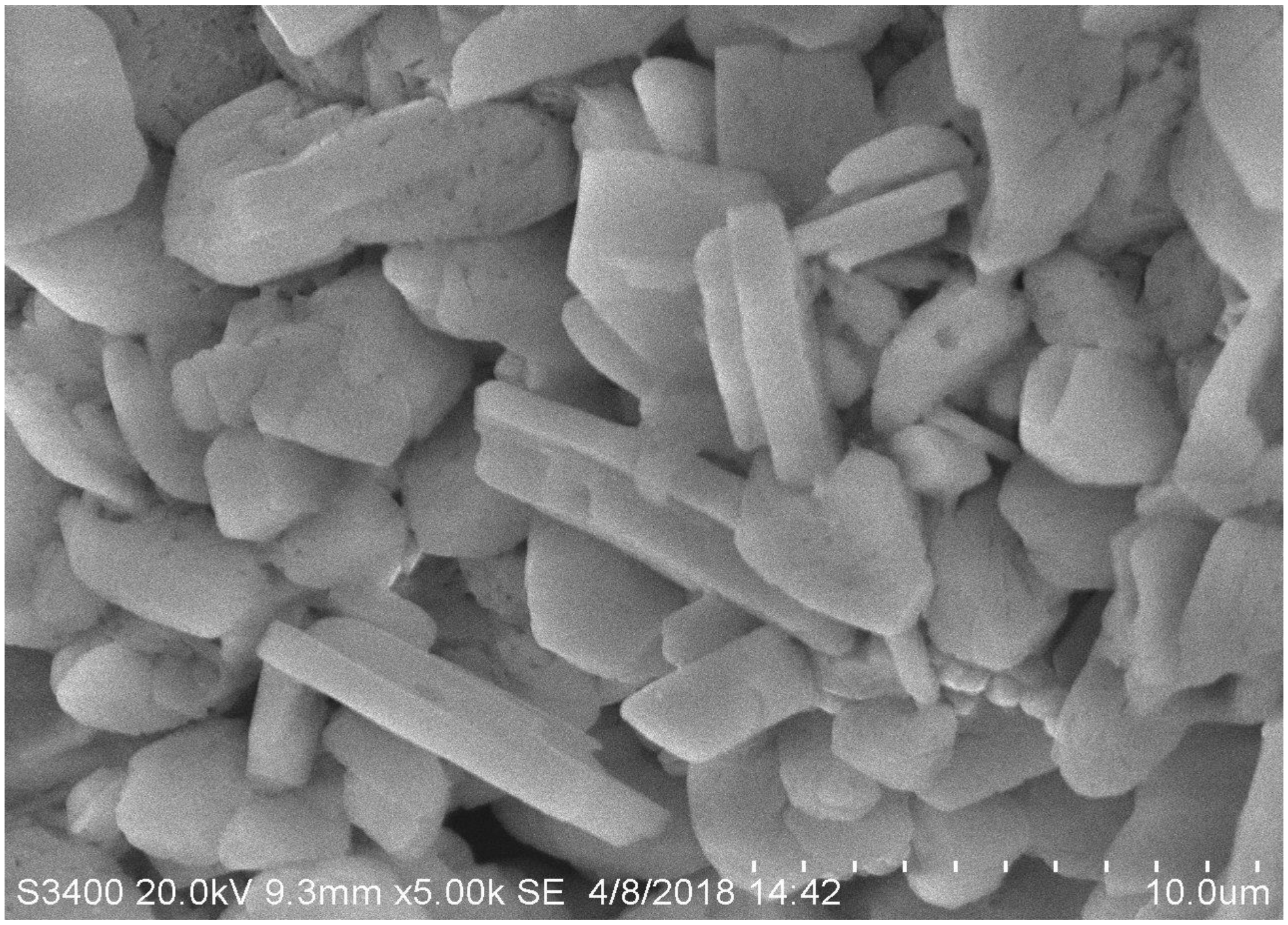

**Figure S3.** SEM image of Sample 1. An exact estimate of the grain size is difficult. Here we take the averaged grain size d as 5 μm according to the SEM image. The bulk sample was soaked with the hydrochloric acid for seconds in order to show the grain boundary clearly.

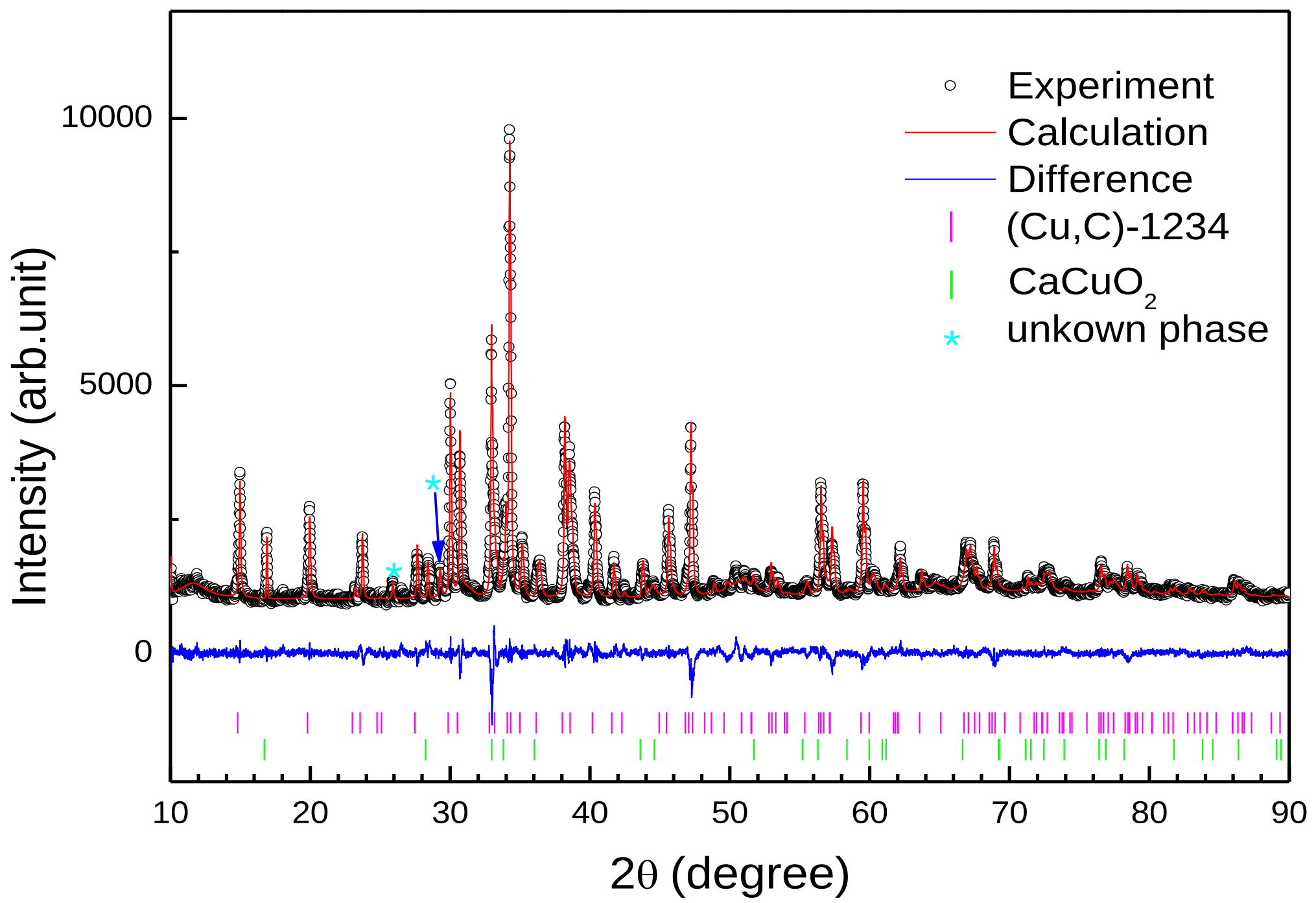

**Figure S4.** Powder x-ray diffraction pattern for Sample 1 (Cu,C)-1234. The XRD can be well indexed by the main phase (Cu,C)-1234 (taking about 93% fraction) and the impurity phase $CaCuO_2$. The main impurity phase $CaCuO_2$ is marked by the vertical grain bars below the spectrum. Two stars mark the unknown phase which takes a very small fraction.